 \documentclass[review,12pt]{elsarticle}

\usepackage[english]{babel}
\usepackage{subfigure}
\usepackage{tabularx}
\usepackage{multirow}
\usepackage{pgf}
\usepackage{tikz}
\usepackage{color}

\usepackage{amssymb}

\usepackage{placeins}

\definecolor{myblue}{rgb}{0.00,0.25,0.50}
\definecolor{mygreen}{rgb}{0.00,0.46,0.00}
\definecolor{myred}{rgb}{.750,0.0,0.00}

\journal{}

\begin{document}
\begin{frontmatter}



\title{3D numerical design of tunnel hood}


\author[label1,label2]{David Uystepruyst\corref{du}}
\cortext[du]{Corresponding author at: TEMPO, Universit\'e de Valenciennes et du Hainaut-Cambr\'esis, 59313 Valenciennes Cedex 9, France.}
\ead{david.uystepruyst@univ-valenciennes.fr}
\author[label3]{Mame William-Louis}
\author[label1,label2]{Fran\c cois Monnoyer}
\address[label1]{Universit\'e Lille Nord-de-France, F-59000 Lille, France.}
\address[label2]{TEMPO, Universit\'e de Valenciennes et du Hainaut-Cambr\'esis, 59313 Valenciennes Cedex 9, France.}
\address[label3]{Univ. Orl\'eans, ENSI de Bourges, PRISME, EA 4229, F45072, Orl\'eans.}

\begin{abstract}

This paper relates to the parametric study of tunnel hoods in order to reduce the shape, i.e the temporal gradient, of the pressure wave generated by the entry of a High speed train in tunnel. This is achieved by using an in-house three-dimensional numerical solver which solves the Eulerian equations on a Cartesian and unstructured mesh.  The efficiency of the numerical methodology is demonstrated through  comparisons with both experimental data and empirical formula. For the tunnel hood design, three parameters, that can influence the wave shape, are considered: the shape, the section and the length of the hood. The numerical results show, (i) that a constant section hood is the most efficient shape when compared to progressive (elliptic or conical) section hoods, (ii) an optimal ratio between hood's section and tunnel section where the temporal gradient of the pressure wave can be reduced by half, (iii) a significant efficiency of the hood's length in the range of 2 to 8 times the length of the train nose. Finally the influence of the train's speed is investigated and results point out that the optimum section is slightly modified by the train's speed. 

\end{abstract}


\end{frontmatter}



\section{Introduction}
\label{sec:intro}

A train entering tunnel generates a compression wave propagating to the opposite portal, where it is partly emitted outside and partly reflected back in the tunnel as a rarefaction wave. The emerging outside part, called the micro-wave, may, in certain circumstances, be strong enough to produce a booming noise up to 140-150~dB. This kind of disturbances can have dramatical effects on the neighborhood environment of the tunnel exit, occurring structural damages and much annoyance.\\

\noindent The magnitude and the duration of those waves are strongly linked to the temporal pressure gradient of the initial compression wave generated by the entry of the train nose in the tunnel.
This temporal pressure gradient can be reduced by modifying the wave generation process. This can be achieved by optimizing the shape of the train nose \cite{mae93,oga_fuj97,bel_kag02,ku10,ki11} or by adding a progressive entry portal to the tunnel \cite{how99,ret_gre02_2,how06,xia10,liu10}. Indeed, the temporal pressure gradient essentially depends on the train Mach number and the blockage ratio, $\sigma$, defined as the ratio of the cross-sectional area of the train to the tunnel entry cross-sectional area. The modifications of the train nose or the addition of a progressive entry allows the evolution of the blockage ratio to be modified and, therefore, the temporal pressure gradient. \\

\noindent Despite the advances in computer efficiency, the parametric study of a transient three-dimensional physical phenomena remains a challenge, and the existing numerical parametric studies are not exhaustive. The oldest study was performed by SNCF researchers \cite{ret_gre02_2}. In this paper, the hood shape and perforations were investigated using only three comparisons. In reference \cite{xia10} the authors compared nine perforated entries having no section discontinuity between hood and tunnel. The optimal hood leads to a reduction of about 43\% of the temporal pressure gradient. The most elaborate study was carried out by Liu \emph{et al.} \cite{liu10}, where a parametric study of the hood section was performed for three hood lengths. The optimal hood section was about 1.8 times the tunnel section for the three hood lengths. Studies of different hood shapes and perforated hoods were carried out as well. However, this study was performed at a low speed of 160~km/h.\\

\noindent The large number of recent works on the optimization of train nose \cite{ku10,ki11}, on the investigation of hoods effects \cite{xia10,liu10,Heine12}, or on simple study of train-tunnel entry \cite{li11,ko12} shows that the problems occurring when trains enter tunnels are still relevant and becomes even more important with the increase of the velocity and the development of high-speed trains in the world.\\

\noindent The main purposes of the present work are to precisely define an optimal hood for a tunnel in which a train is entering at high-speed velocities, higher than 250~km/h, and to thoughtfully describe physical phenomena involved by different hoods. In order to achieve this, a numerical parametric study of the hood was carried out. The considered parameters were the shape, the section and the length of the hood. This introductory part is followed by a description of the numerical model in section \ref{sec:nummod}. This is followed by a description of the geometrical models used in this study in section \ref{ssec:geom}. The capability of the numerical method to correctly reproduce the phenomena is shown in section \ref{sec:val}.  Afterwards, the hood geometries and the results of the studies of the shape, the section and the length of the hood are provided in section \ref{ssec:shape},  \ref{ssec:section} and  \ref{ssec:length}, respectively. In section  \ref{ssec:velocity}, the effect of the train velocity is studied. 

\section{The numerical model}
\label{sec:nummod}

\subsection{Governing equations}

When a train enters a tunnel, the pressure forces are strongly dominant as compared to the viscous forces. Therefore, the viscosity was neglected as well as the turbulence of the flow. The simulations were modeled by the three-dimensional equations of conservation of mass, quantity of movement and energy:

\begin{equation}
    \partial_{t}\mathbf{U}+\nabla\cdot\mathbf{H}(\mathbf{U})=0,
    \label{equa1}
\end{equation}

\noindent where $\mathbf{U}$ is the vector of conservative variable and $\mathbf{H}(\mathbf{U})$ is the fluxes tensor :
$$
    \mathbf{U}=\left(
    \begin{array}{c}
    \rho \\
    \rho u\\
    \rho v \\
    \rho w \\
    \rho E
    \end{array} \right),
    \textmd{ and }    
    \mathbf{H}(\mathbf{U})=\left(
    \begin{array}{ccc}
    \rho u_{r} & \rho v_{r} & \rho w_{r} \\
    \rho uu_{r}+p & \rho uv_{r} & \rho uw_{r} \\
    \rho vu_{r} & \rho vv_{r}+p & \rho vw_{r}\\
    \rho wu_{r} & \rho wv_{r} & \rho ww_{r}+p\\
    \rho u_{r}h_{0}+pu_{t} & \rho v_{r}h_{0}+pv_{t} & \rho w_{r}h_{0}+pw_{t}
    \end{array} \right).
$$

\noindent In the definitions of $\mathbf{U}$ and $\mathbf{H}(\mathbf{U})$, the variables are the density $\rho$, the energy $E$, the pressure $p$ and the total enthalpy $h_0$. $u_r$, $v_r$ and $w_r$ are the three components of the relative velocity $\mathbf{V}_r$, defined as $\mathbf{V}_r=\mathbf{V}-\mathbf{V}_t$ where $\mathbf{V}=(u,v,w)^t$ is the absolute velocity of the flow and $\mathbf{V}_t=(u_t,v_t,w_t)^t$ is the translation velocity of the grid.\\

\noindent Equation (\ref{equa1}) was solved using a finite volume method. It combines the second order Roe scheme \cite{roe81} for the spatial discretization and the Van Leer predictor-corrector scheme \cite{van79} for the time integration. To preserve the monotonous property of the scheme during the passage at the second order, the limiter for non-structured mesh of type Barth and Jespersen \cite{bar_jes89} was used.

\subsection{The numerical domain and boundary conditions}

\begin{figure}[!h]
\centering
\begin{tikzpicture}[set style={{help lines}+=[dashed]},scale=.07]
\draw [very thick] (120,30) -- (130,30) node [right] {solid wall};
\draw (120,25) -- (130,25) node [right] {external domain};
\draw [mygreen] (120,20) -- (130,20) node [right] {symmetry};
\draw [myblue, dashed] (120,15) -- (130,15) node [right] {sliding domain};
\draw (0,2) -- (0,15) -- (100,15); \draw[very thick] (100,15) -- (100,4.514) --
(200,4.514); \draw (200,4.514) -- (200,2);
\filldraw[fill=gray!20,thick,rounded corners=2pt] (5.5,0) -- (6,1.4) -- (26.48,1.4) -- (26.5,0);
\filldraw[fill=gray!20,thick,rounded corners=2pt] (56.5,0) -- (57,1.4) -- (77.48,1.4) -- (77.5,0);
\draw[myblue, dashed] (-5,0) -- (-5,2) -- (205,2) -- (205,0);
\draw[mygreen] (-5,0) -- (205,0);
\draw [dashed] (5,20) -- (5,-3); \draw [dashed] (26.48,20) -- (26.48,-3); \node at (15.74,21) {\small initial};  \node at (15.74,17) {\small position};
\draw [dashed] (56,20) -- (56,0); \draw [dashed] (77.48,20) -- (77.48,0); \node at (66.74,21) {\small restart};  \node at (66.74,17) {\small position};
\draw[->] (15.74,-3) -- (5,-3); \draw[->] (15.74,-3) -- (26.48,-3); \node
at (15.74,-6) {\textmd{\small$96~m$}}; \draw [dashed] (100,0) --
(100,-3); \draw[->] (63.24,-3) -- (100,-3); \draw[->] (63.24,-3)
-- (26.48,-3); \node at (63.24,-6) {\textmd{$190~m$}};
 \draw [dashed] (0,0) -- (0,-12); \draw [dashed]
(100,0) -- (100,-12); \draw[->] (50,-12) -- (0,-12); \draw[->]
(50,-12) -- (100,-12); \node at (50,-30) {\textmd{$300~m$}}; \draw
[dashed] (200,-4) -- (200,-12); \draw[->] (150,-12) -- (100,-12);
\draw[->] (150,-12) -- (200,-12); \node at (150,-15)
{\textmd{$300~m$}}; \draw [dashed] (0,15) -- (-10,15); \draw
[dashed] (0,0) -- (-10,0); \draw [->] (-10,5) -- (-10,15); \draw
[->] (-10,5) -- (-10,0); \node at (-20,7.5) {\textmd{$80~m$}};
\filldraw (133,4.514) circle (10pt) node [above] {numerical gauge};
\draw[dashed] (133,4.514) -- (133,-3); \draw[->] (115,-3) -- (100,-3); \draw[->] (115,-3) -- (133,-3); \node at (116.5,-6) {\textmd{$100~m$}};
\end{tikzpicture}
\caption{\emph{Numerical domain}.}\label{param_dim}
\end{figure}
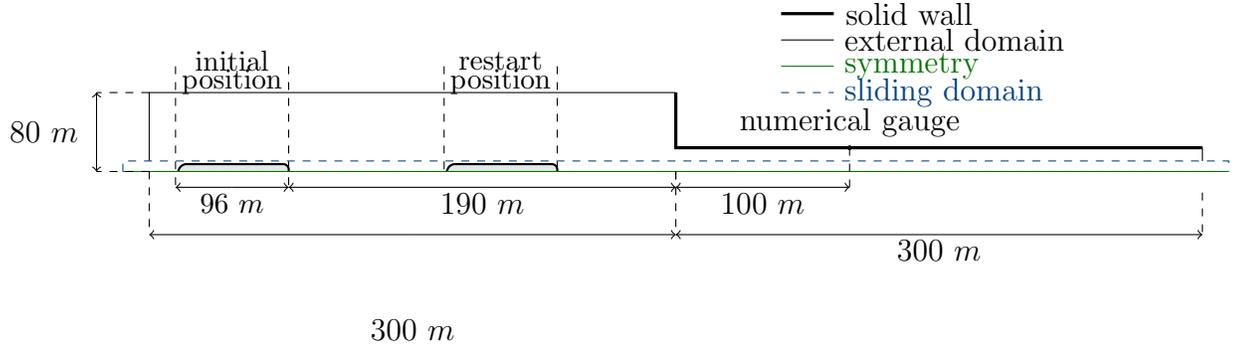

A top view of the numerical domain is represented in figure \ref{param_dim}. The overall domain was subdivided in two sub-domains. The first one, the sliding domain, contained the train and was set in motion with the translation velocity $\mathbf{V}_t$. The remaining part of the overall domain contained the tunnel walls and the external domain. This second part stayed motionless during simulations, the translation velocity $\mathbf{V}_t$ of equation (\ref{equa1}) was, hence, equal to zero for its computational cells.\\

\noindent A solid walls boundary condition was used for the tunnel walls and for the train body. A non reflective boundary condition, based on a Fourier decomposition of the solution at the boundary face \cite{uys11}, was used for the external domain and the tunnel extremity. The cross-sectional section of the train was centered inside the tunnel, a symmetry boundary condition was then used to divide the computational domain by two. A common interface was calculated between the sliding domain and the second domain, and a conservative flux calculation was performed, i.e. at a common interface the flow variables were updated by calculating a flux and not by interpolating, see \cite{uys11} for more details. 

\subsection{Geometrical models and numerical details}
\label{ssec:geom}

\begin{figure}[h!]
\centering
\subfigure[Nose of the TGV]{
\includegraphics[scale=.35]{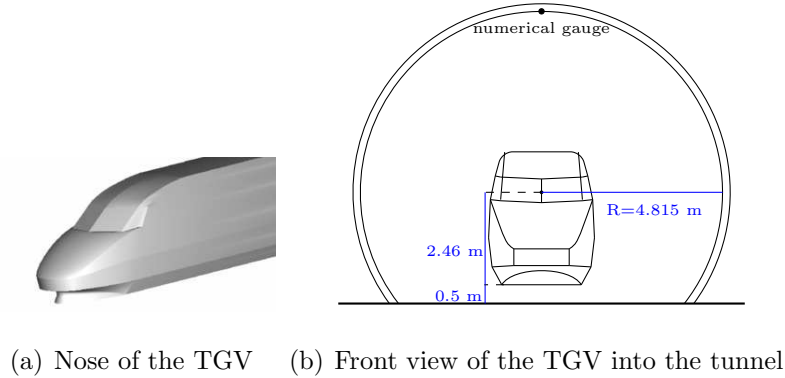}}
\subfigure[Front view of the TGV into the tunnel]{
\begin{tikzpicture}[set style={{help lines}+=[dashed]},scale=0.5]
\draw (3.797,0) arc (-37.9:217.9:4.815); \draw (4.0483,0)
arc(-36.2:216.2:5.015); \draw[thick] (-5.404,0) -- (5.404,0);
\draw[very thin,white] (-6.5,0) -- (-6.5001,0); \draw[very thin,white] (6.5,0) -- (6.5001,0);
\filldraw [] (0,2.96) circle (1pt); \filldraw [] (0,7.775) circle
(2pt) node [below] {\tiny\textmd{numerical gauge}};
\draw (-1.057,0.5) -- (1.057,0.5); \draw (-1.057,0.5) --
(-1.316,0.975); \draw (1.057,0.5) -- (1.316,0.975); \draw
(-1.407,1.747) -- (-1.316,0.975); \draw (1.407,1.747) --
(1.316,0.975);  \draw (-1.407,1.747) -- (-1.36,2.765);  \draw
(1.407,1.747) -- (1.36,2.765); \draw (-1.23,3.395) -- (-1.36,2.765);
\draw (1.23,3.395) -- (1.36,2.765); \draw (-1.23,3.395) {[rounded
corners] -- (-1.052,4.0369) -- (1.052,4.0369)} -- (1.23,3.395);
\draw (1.087,2.765) -- (0.98,4.03); \draw (-1.087,2.765) --
(-0.98,4.03); \draw (1.36,2.765) -- (1.087,2.735) -- (0,2.685) --
(-1.087,2.735) -- (-1.36,2.765); \draw (1.087,3.365) -- (0,3.315) --
 (-1.087,3.365); \draw (0,3.315) -- (0,2.685); \draw
(1.23,3.395) -- (1.087,3.365); \draw (-1.23,3.395) --
(-1.087,3.365); \draw (-1.316,0.975) -- (-0.736,1.01); \draw
(1.316,0.975) -- (0.736,1.01); \draw (-0.736,1.01) -- (0.736,1.01);
\draw (0.736,1.455) -- (-0.736,1.455); \draw (0.736,1.455) --
(0.736,1.01); \draw (-0.736,1.455) -- (-0.736,1.01); \draw
(-1.057,0.5) .. controls (-0.75,1.0) and (0.75,1.0) .. (1.057,0.5);
\draw (-1.36,2.765) .. controls (-0.95,1.6) .. (-0.736,1.455); \draw
(1.36,2.765) .. controls (0.95,1.6) .. (0.736,1.455);
\draw[blue] (0,2.96) -- (4.815,2.96) node at (3,2.5) {\tiny R=4.815~m};
\draw[blue] (-1.5,0) -- (-1.5,0.5) node at (-2.2, 0.2) {\tiny 0.5~m};
\draw[dashed] (-1,0.5) -- (-1.52,0.5); \draw[blue] (-1.5,2.96) --
(-1.5,0.5) node at (-2.3,1.405) {\tiny 2.46~m}; \draw[dashed] (0,2.96)
-- (-1.52,2.96);
\end{tikzpicture}}
\caption{\emph{Geometric properties (in m).}}
\label{fig:geomprop}
\end{figure}

For this study, the entry of the French TGV, see figure , into a $63~m^{2}$ section area tunnel at a speed of 250~km/h ($M\simeq0.2$ with $T=293~K$), except for section \ref{ssec:velocity}, was considered. The geometries of the TGV and the tunnel are shown in figure \ref{fig:geomprop}. The blockage ratio $\sigma$ was $0.152$.\\

\noindent The train length was $96~m$, hence the rarefaction wave generated by the rear train entry did not affect the pressure signal recorded by a numerical sensor located $100~m$ after the tunnel entry. \\

\noindent The computational grid was performed by an automatic grid generator which made Cartesian volume mesh based on a triangular surface mesh. The grid generator used an Octree structure which means, in particular, that the size ratio between two neighbor computational cells cannot be different to one or two. A cut plane view of the volume mesh is shown in figure \ref{vmesh} for the configuration without hood and for one configuration with hood. The overall domain contained between 1.250.000 and 1.350.000 elements. 
The minimum mesh size was located on the train nose, as can be seen in both figures \ref{vmesh_h0} and \ref{vmesh_h1}, and was about $0.05~m$. For the train sub-domain, i.e. the sliding sub-domain in figure \ref{param_dim}, the mesh size grew to $0.1~m$, and then $0.2~m$ in front of the train. A fine mesh was required in front of the train for the good simulation of the compression wave. Others sub-domains were finely meshed into the hood and into the tunnel, as shown in figures \ref{vmesh_h0} and \ref{vmesh_h1}. where the minimal space discretization is $0.2~m$. The time-step was $\Delta t=3.4\times10^{-5}~s$.

\section{Validation}
\label{sec:val}

\subsection{Tunnel without hood}
\label{ssec:compform}

First, the reference configuration described in section \ref{ssec:geom} had to be validated. Unfortunately, experimental data were not available for this configuration. Nevertheless, it was possible to determine the maximum value of pressure, as well as the maximum value of pressure gradient, with empirical formula. These formula depend on parameters which can be determined by experimental measurements. Some previous experimental measurements were available such the case of the TGV running in the Villejuste tunnel \cite{ret_gre02_1}. As soon as the parameters were determined, it was possible to estimate the maximum values of both pressure and pressure gradient for the reference configuration of the present parametric study.\\

\noindent The Villejuste tunnel located on the high-speed southwest line in France, has a section of $46~m^2$, giving a blockage ratio of $0.21$. The velocity of the TGV  was $220~km/h$ \cite{ret_gre02_1}.\\

\noindent The experimental data and the numerical results of the pressure rise and the maximum of pressure gradient are presented in table \ref{max_vil}. It is shown that the numerical methodology gives results in good convenience with experimental data.\\

\begin{table}[!t]
\begin{center}
\begin{tabular}{|c||c|c||c|c|}
\hline  & \multicolumn{2}{c||}{$\Delta P$ max} &
\multicolumn{2}{c|}{$\partial P/\partial t$ max} \\
\hline & value (Pa) & diff./ref. (\%) & value
(Pa/s) &
diff./ref. (\%) \\
\hline
Experimental & 1281 &  & 8400 &  \\
 \hline
Numerical & 1251 & -2 & 8795 & +4 \\
  \hline
\end{tabular}
\end{center}
 \caption{\emph{Experimental data and numerical results of the maxima of pressure and pressure gradient for the entrance of the TGV in the Villejuste tunnel \cite{ret_gre02_1}. $V=220~km/h$, $\sigma=0.21$.}}
 \label{max_vil}
\end{table}

\noindent These maxima can be also estimated by formula developed by Pope \cite{woo_pop92}, also presented in \cite{mwl_tou05}, for the pressure rise:

\begin{equation}
    \Delta p=\gamma p_{0}M\left[1+\displaystyle\frac{1-\sqrt{1+2Y}}{Y}\right],
    \label{dp_equa}
\end{equation}
where $Y=Mk_{n}(R^{2}-1)$ with $k_{n}=1+\displaystyle\frac{\xi_{n}}{1-(1-\sigma)^{2}}$ and
$R=\displaystyle\frac{1}{1-\sigma}$. $M$ is the train Mach number, $p_{0}$ is the free pressure and $\xi_{n}$ is pressure loss coefficient of the train nose;\\
\noindent Or by the formula of Ozawa \cite{oza_mae88} for the pressure and the pressure gradient:

\begin{equation}
    \Delta p=\displaystyle\frac{\rho}{2}V^{2}f(M,\sigma),
    \label{p_max_formula}
\end{equation}

\begin{equation}
    \left(\partial_t p\right)_{max}=\eta\displaystyle\frac{\rho}{2}V^{3}f(M,\sigma)\displaystyle\frac{1}{d_{tun}},
    \label{grad_max_formula}
\end{equation}
with
\[
 f(M,\sigma)=\displaystyle\frac{1-(1-\sigma)^{2}}{\left(1-M\right)\left(M+(1-\sigma)^{2}\right)},
\]
where $\rho$ is the density, $V$ is the train velocity, $d_{tun}$ is the hydraulic diameter of the tunnel and $\eta$ is a coefficient depending of the train and the tunnel.\\

\noindent For the following parametric study, the pressure loss coefficient of the nose train of equation (\ref{dp_equa}) can be estimated from the experimental value of $\Delta p$ given in table \ref{max_vil}. It gives a value of $0.02$ for the pressure loss coefficient of the TGV. In the same way, the coefficient $\eta$ of equation (\ref{grad_max_formula}) can be approximated to $0.82$ with the experimental value of pressure gradient maximum. The knowledge of these coefficients allow to apply the formula (\ref{dp_equa}), (\ref{p_max_formula}) and (\ref{grad_max_formula}) to the configuration described in section \ref{ssec:geom}, and used in the following parametric study. The values given by the formula and the numerical results are compared in table \ref{formula_num}, for the pressure and the pressure gradient.\\

\begin{table}[!t]
\begin{center}
\begin{tabular}{|c||c|c|c|}
\hline 
 & Ozawa & Pope & Numerical\\
\hline
$\Delta P$ (Pa) & 1113 & 1126 & 1144 \\
\hline
$\partial_t P$ max (Pa/s) & 7082 & & 7522 \\
\hline
\end{tabular}
\end{center}
 \caption{\emph{Formula values and numerical results of the maxima of pressure and pressure gradient for the entrance of the TGV in the $63~m^2$ section area tunnel. $V=250~km/h$, $\sigma=0.152$}.}
 \label{formula_num}
\end{table}

\noindent The numerical values are in good agreement with the formula values. The difference is only 3\% for the pressure rise and 6\% for the maximum of pressure gradient. As said previously, the coefficient $\eta$ depends especially on the tunnel entry and can be slightly different for the $63~m^2$ tunnel compared to the value obtained for the Villejuste tunnel. This slight discrepancy can explain the low value of pressure gradient given by the formula (\ref{grad_max_formula}).

\subsection{Tunnel with hood}
\label{ssec:compexp}

\noindent Here, comparisons are performed with the experimental data obtained by Bellenoue et al \cite{bel01}. In this experimental work, a reduced scale train of $600~mm$ in length was used. This train was a cylinder of $25~mm$ in diameter with an elliptic shape nose of $40~mm$ in length and a flat tail. The tunnel, as the hood, was a cylinder. The inner diameter of the tunnel was $44~mm$ giving a blockage ratio train/tunnel of $\sigma=0.32$. In \cite{bel01}, several hood sections, as well as several hood lengths were studied. For the present comparison, the selected hood was such that the ratio between the hood section and the tunnel section was $S_h/S_{tun}=1.7$, and its length was four times the length of the train nose $L_h=4L_{nose}$. The velocity of the train was $43~m.s^{-1}$.

\begin{figure}[!h]
\centering \subfigure[pressure]{
\includegraphics[width=0.45\textwidth]{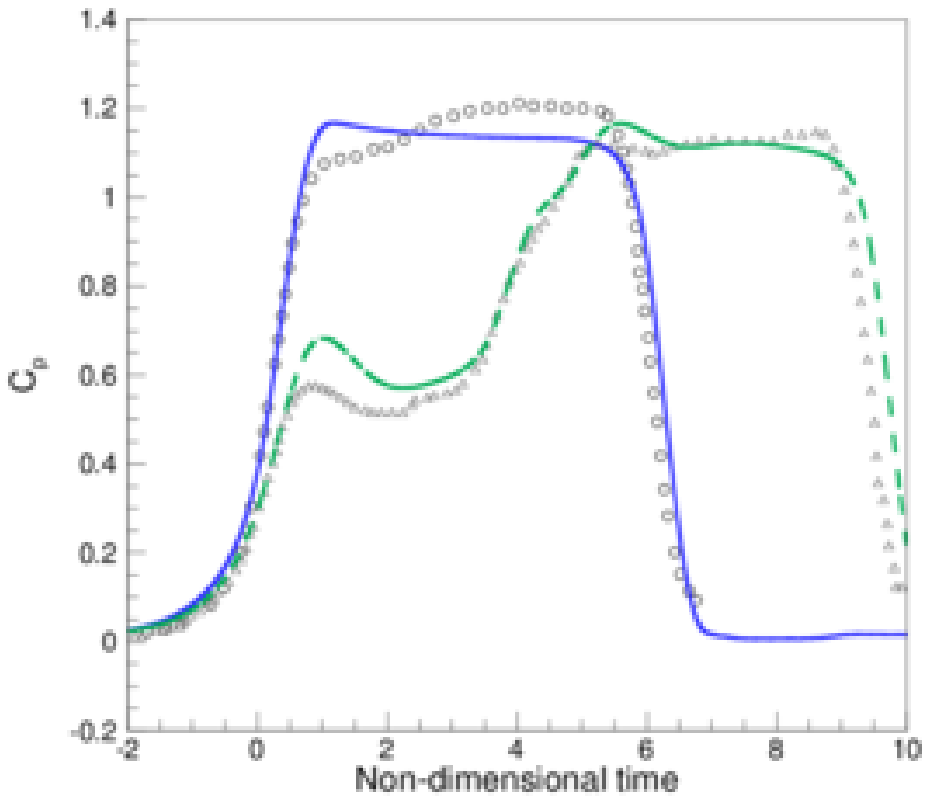}}
\subfigure[pressure gradient]{
\includegraphics[width=0.45\textwidth]{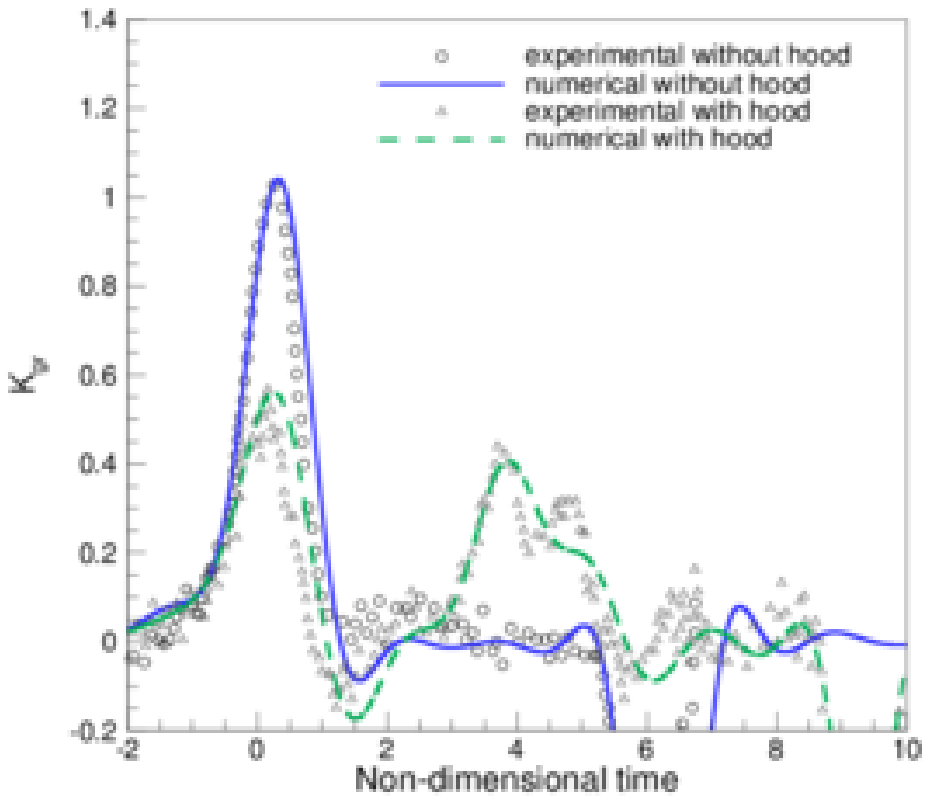}}
\caption{\emph{Temporal history of the pressure and its gradient. Experimental data from \cite{bel01}, and numerical results. $V=43~m.s{-1}$, $\sigma=0.32$, $S_h/S_{tun}=1.7$ and $L_h=4L_{nose}$.}}\label{saut_bel}
\end{figure}

\noindent Figure \ref{saut_bel} shows the comparisons, for the pressure and the pressure gradient, between the experimental data of \cite{bel01} and the numerical results. For convenience, the pressure, the pressure gradient and the time are presented by non-dimensional coefficients, as in the experimental work. The non-dimensional pressure $C_p$, pressure gradient $K_{gr}$ and time $t_a$ are, then, given by:
\[
 C_p=\displaystyle\frac{p-p_0}{\frac{1}{2}\rho V^2},~ K_{gr}=\displaystyle\frac{\partial_t p}{\displaystyle\frac{1}{2}\rho\frac{V^3}{d_{tun}}}\textmd{ and }t_a=\displaystyle\frac{V(t-t_c)}{L_{nose}}\textmd{ with }t_c=(L_h+X_{sensor})/c,
\]
where $c$ is the sound speed, $t_c$ is then the time needed by the pressure wave to reach the gauge located at distance $X_{nose}$ from the tunnel entrance.\\

\noindent It is shown that the convenience between both approaches is rather good. The maxima values of the pressure gradients are, especially, well determined by the numerical methodology. Moreover, it is shown that the numerical methodology is suitable to reproduce main phenomena : compression waves, rarefaction wave, etc.

\section{Results}
\label{sec:results}

\subsection{Hood shape}
\label{ssec:shape}

\begin{figure}[!t]
\centering
\begin{tikzpicture}[set style={{help lines}+=[dashed]},scale=.07]
\draw (-10,60) -- (50,60); \draw (40,90) -- (40,75) -- (50,75); \draw (40,62) -- (40,58) node [below] {$x=0$}; \draw[->,dashed] (45,67) -- (45,60); \draw[->,dashed] (45,67) -- (45,75); \node at (39,67.5) {$S_{tun}$}; \node at (20,48) {without hood};
\draw (120,60) -- (180,60); \draw (130,90) -- (170,90) -- (170,75) -- (180,75); \draw (170,62) -- (170,58) node [below] {$x=0$}; \draw (130,62) -- (130,58) node [below] {$x=-L_{h}$}; \draw[->,dashed] (130,75) -- (130,90); \draw[->,dashed] (130,75) -- (130,60); \node at (125,75) {$S_{h}$}; \draw[->,dashed] (175,67) -- (175,60); \draw[->,dashed] (175,67) -- (175,75); \node at (169,67.5) {$S_{tun}$}; \node at (150,48) {constant section hood};
\draw (-10,0) -- (50,0); \draw (0,30) -- (40,15) -- (50,15); \draw (40,2) -- (40,-2) node [below] {$x=0$}; \draw (0,2) -- (0,-2) node [below] {$x=-L_{h}$}; \draw[->,dashed] (0,15) -- (0,30); \draw[->,dashed] (0,15) -- (0,0); \node at (5,15) {$S_{h}$}; \draw[->,dashed] (45,7) -- (45,0); \draw[->,dashed] (45,7) -- (45,15); \node at (39,7.5) {$S_{tun}$}; \node at (20,-12) {conical progressive hood};
\draw (120,0) -- (180,0); \draw (170,15) arc (-90:-180:40 and
15); \draw (170,15) -- (180,15); \draw (170,2) -- (170,-2) node [below] {$x=0$}; \draw (130,2) -- (130,-2) node [below] {$x=-L_{h}$}; \draw[->,dashed] (130,15) -- (130,30); \draw[->,dashed] (130,15) -- (130,0); \node at (125,15) {$S_{h}$}; \draw[->,dashed] (175,7) -- (175,0); \draw[->,dashed] (175,7) -- (175,15); \node at (169,7.5) {$S_{tun}$}; \node at (150,-12) {elliptic progressive hood};
\draw (92.594,45) arc (-37.93:217.93:9.63); \draw[thick] (65,45)
-- (105,45); \filldraw (85,50.92) circle(1pt); \draw (85,50.92) --
(87,60.34); \node at (82.4,56)
{\tiny\textmd{$R_{tun}$}}; \draw
(85,50.92) -- (68,50.92); \draw [->] (68,47.96) -- (68,45); \draw [->]
(68,47.96) -- (68,50.92); \node at (64.8,47.96) {\tiny\textmd{$h$}};
\draw[blue] (97.2638,45) arc (-25.75:205.75:13.618); \draw[blue] (85,50.92) --
(98.456,53); \node at (103,53.6)
{\textcolor[rgb]{0.00,0.00,1.00}{\tiny\textmd{$R_{h}$}}};
\node at (85,41) {longitudinal cross section};
\node at (85,84) {$h=2.96~m$};
\node at (85,77) {$R_{tun}=4.815~m$};
\node at (85,70) {$R_{h}=6.809~m$};
\end{tikzpicture}
\caption{\emph{Hood shape.}}\label{3auvents}
\end{figure}
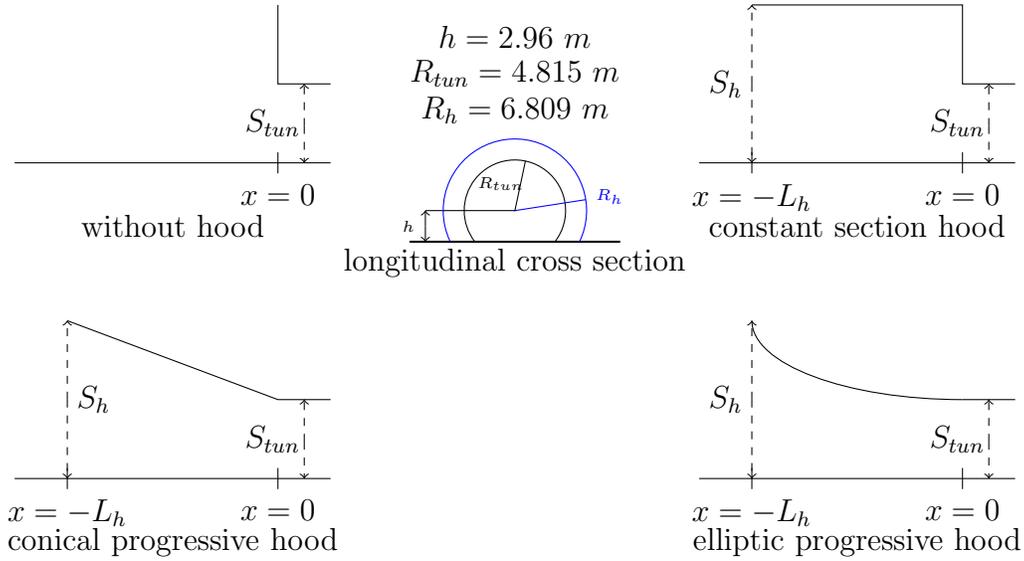

\subsubsection{The test layout}

In order to highlight the effect of the hood shape, four cases were simulated, see figure \ref{3auvents}.
The first one concerned a tunnel without a hood, and is considered as the reference for all other calculations. The second one was a hood with a constant section, and the two last were progressive hoods with a section progressively reducing to that of the tunnel. This progressive evolution was conical for the third configuration and elliptical for the last. For this study, the hood length, $L_{h}$, was fixed at 20~m and the hood section, $S_{h}$, is 1.775 times the tunnel section, $S_{tun}$. The tunnel and hood cross sections are two concentric circles truncated by the ground with hood radius defined as $R_{h}=\sqrt{2}R_{tun}$, see figure \ref{3auvents}.

\subsubsection{Results}

\begin{figure}[!h]
\centering \subfigure[pressure]{
\includegraphics[width=0.45\textwidth]{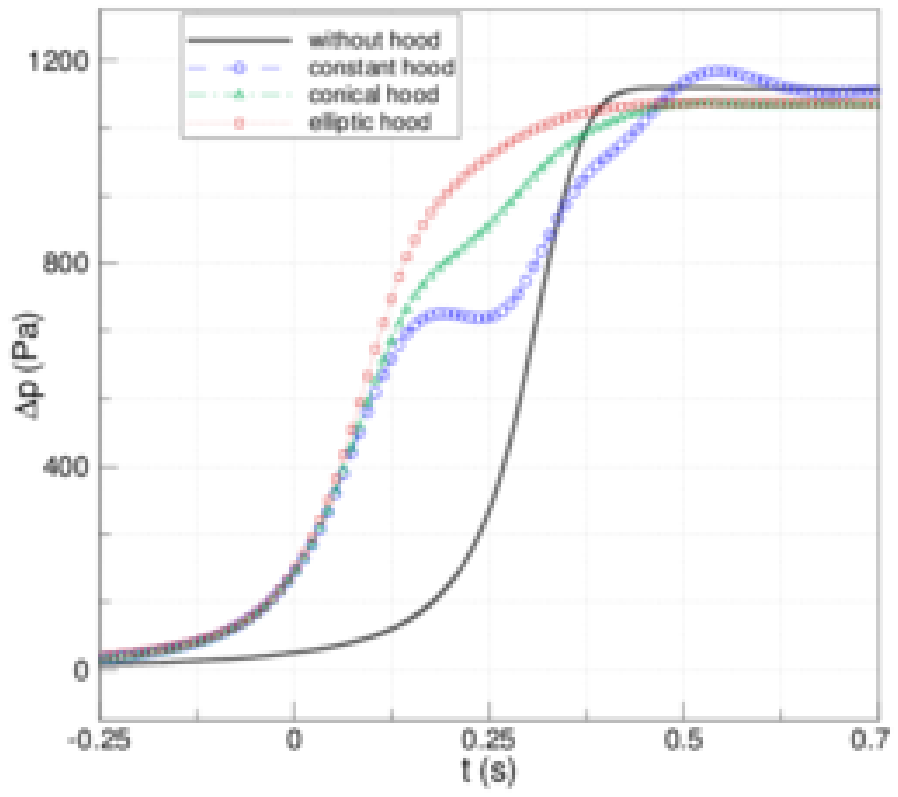}}
\subfigure[pressure gradient]{
\includegraphics[width=0.45\textwidth]{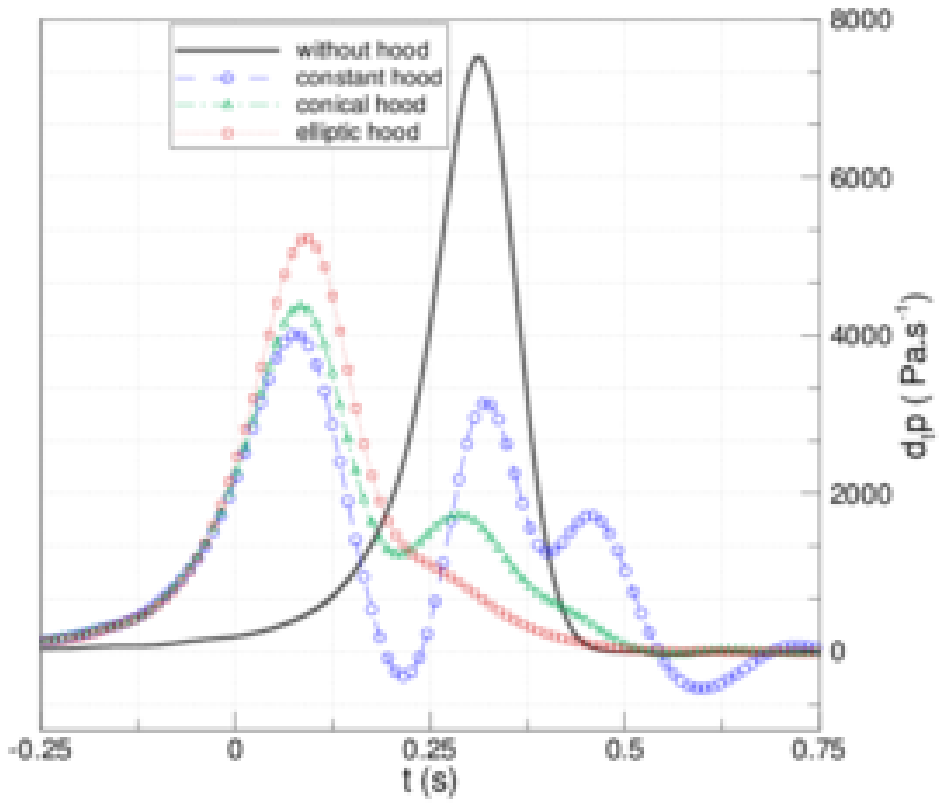}}
\caption{\emph{Temporal history of the pressure and its gradient. Numerical results obtained on the four hood shapes defined in \ref{3auvents}. $V=250~km/h$, $\sigma=0.152$, $S_h/S_{tun}=1.775$ and $L_h=20~m$.}}\label{saut_param1}
\end{figure}

Figure \ref{saut_param1} shows the temporal evolution of the pressure, recorded at 100~m from the tunnel entry, and its temporal gradient. On these graphs, the time $t=0~s$ is the train/tunnel entry instant. \\

\noindent The configuration without a hood clearly produces a single jump in pressure. It is also clear that configurations with hoods considerably reduce the pressure gradient. Both progressive hoods imply a substantial pressure increase at first which corresponds to the entry of the train in the hood. This substantial increase leads to a first peak of gradient. During a second phase, the pressure increases slowly due to the narrowing of the hood section and, therefore the rise of the train-hood blockage ratio.\\

\noindent The elliptic hood yield an important first pressure jump, involving a pressure gradient maximum of $5326~Pa/s$, see table \ref{max_param1}. The decrease of its cross-sectional area was faster than that of the conical hood, see figure \ref{3auvents}. That means at the same time, the blockage ratio with the conical hood was, then, higher and, finally, the pressure, as well as the pressure gradient, was higher. For the conical hood, the changes in cross-sectional area were more important at the end of the hood, this yields a second pressure jump before the train entered the tunnel around time $t=0.25~s$, see figure \ref{saut_param1}. As the maximum value of pressure gradient was $4368~Pa/s$, it can be concluded that the conical hood was more efficient than the elliptical hood. However, as it was shown by Ogawa and Fujii \cite{oga_fuj96} for the train noses, a hood with shape made by a combination of elliptical, conical or even paraboloidal shape could be more efficient.\\

\begin{figure}[!h]
\centering
\includegraphics[bb=105 30 695 560,clip=true,scale=.5]{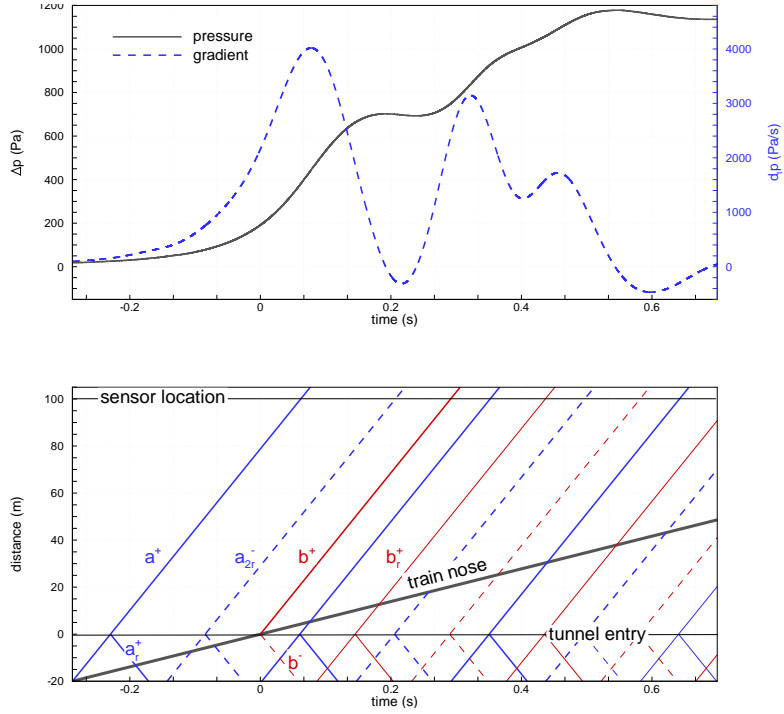}
\caption{\emph{Wave diagram of the constant hood. $V=250~km/h$, $\sigma=0.152$, $S_h/S_{tun}=1.775$ and $L_h=20~m$.}}\label{diag_wave}
\end{figure}

\noindent The constant hood generates a more complex pressure signal. In order to have a better understanding of this pressure evolution, reference may be made to the wave diagram in figure \ref{diag_wave}.\\

\noindent The \emph{x-axis} of this diagram (bottom of figure \ref{diag_wave}) is time, and is the same range as the pressure and gradient signals (top of figure \ref{diag_wave}). The time $t=0~s$ corresponds to the train's entry in the tunnel. The $y-axis$ corresponds to the distance traveled in the hood and the tunnel. The distance $d=-20~m$ is the hood entry, $d=0~m$ is the tunnel entry and $d=100~m$ is the location of the pressure gauge. Finally, the slanting line is the train nose path inside the hood and the tunnel.\\

\begin{table}[!t]
\begin{center}
\begin{tabular}{|c||c|c||c|c|}
\hline  & \multicolumn{2}{c||}{$\Delta P$ max} &
\multicolumn{2}{c|}{$\partial P/\partial t$ max} \\
\hline & value (Pa) & diff./ref. (\%) & value
(Pa/s) &
diff./ref. (\%) \\
\hline
without hood & 1144 &  & 7522 &  \\
 \hline
constant section hood & 1177 & +3 & 4018 & -46 \\
  \hline
conical progressive hood & 1113 & -3 & 4368 & -42 \\
  \hline
elliptical progressive hood & 1116 & -3 & 5326 & -30 \\
  \hline
\end{tabular}
\end{center}
 \caption{\emph{Shape hood maxima pressure amplitude and gradient. Numerical results obtained on the four hood shapes defined in \ref{3auvents}. $V=250~km/h$, $\sigma=0.152$, $S_h/S_{tun}=1.775$ and $L_h=20~m$.}}
 \label{max_param1}
\end{table}

\noindent When the train enters the hood ($t\simeq-0.288~s$ and $d=-20~m$), it generates a compression wave $a^{+}$ which reaches the sensor at time $t=0.05~s$, thereby generating the first peak. Before that, this wave arrives at the tunnel entry where it is partly reflected, producing compression wave $a^{+}_{r}$, on the wall corresponding to the section discontinuity between hood and tunnel. Wave $a^{+}_{r}$ propagates through the hood back towards its entry where it is partly emitted outside and partly reflected back into the hood as rarefaction wave $a^{-}_{2r}$. This reflection is not instantaneous due to the generation of transversal waves at corners \cite{mwl_tou03}. Therefore, the reflection delay can be approximated by the time that the sound travels the hydraulic diameter of the section defined as the difference between hood section and train section. This rarefaction wave reaches the sensor just after time $t=0.2~s$, inducing the pressure decrease and the negative gradient. \\

\noindent When the train enters the tunnel ($t=0~s$ and $d=0~m$), it generates a compression wave $b^{+}$ inside the tunnel and a rarefaction wave $b^{-}$ inside the hood \cite{iid01}. The compression wave generates the second pressure jump and is immediately followed by another compression wave $b^{+}_{r}$ resulting from the reflection of $b^{-}$ at the hood entry. The third jump, caused by $b^{+}_{r}$, explains the overestimation of the pressure amplitude generated by the constant section hood in comparison to the reference case.\\

\noindent The elliptic progressive hood gives a reduction of about only 30\% on the pressure gradient (Table \ref{max_param1}), while with the conical progressive hood, a gain of 42\% is obtained. The most efficient hood is the one with a constant section, which leads to a 46\% reduction. These last two hoods have quasi equivalent results. However, constant section hood implies a more sophisticated pressure signal and, especially, it allow a rarefaction wave to pass between both main compression waves. For this reason the constant section hood is considered for the following studies.\\


\noindent It is clear that the hood section is primordial in determining the intensity of the first jump in pressure gradient. Namely, a small hood section produces a substantial first jump and, in contrast, a large hood section implies a small first jump. This first jump allows the air inside the tunnel to be made to move. This flow velocity is proportional to the intensity of the first jump, and determines the intensity of the second jump. Therefore, a small hood section leads to a high air velocity in the tunnel and it implies a second jump of low intensity. On the contrary, a large hood section generates low air velocity and substantial second jump. In order to obtain the optimal section leading to the "equalization" of both pressure gradient maxima, a parametric study of the hood section was performed.

\subsection{Hood section}\label{hs}
\label{ssec:section}

\subsubsection{The test layout}

In order to determine the optimal hood section, 11 calculations were performed with different sections.
Hood sections ranged from the smallest R1, with a hood/tunnel ratio of 1.397, to the largest R11, with a hood/tunnel ratio of 3.203. The ratio hood/tunnel is indicated in table \ref{max_param2} for each configuration. Case R0 taken as the reference consists of a tunnel without a hood.

\subsubsection{Results}

Figure \ref{saut_param1} shows the temporal evolution of the pressure, recorded at the numerical sensor, and its gradient.\\

\begin{figure}[!h]
\centering \subfigure[pressure]{
\includegraphics[width=0.45\textwidth]{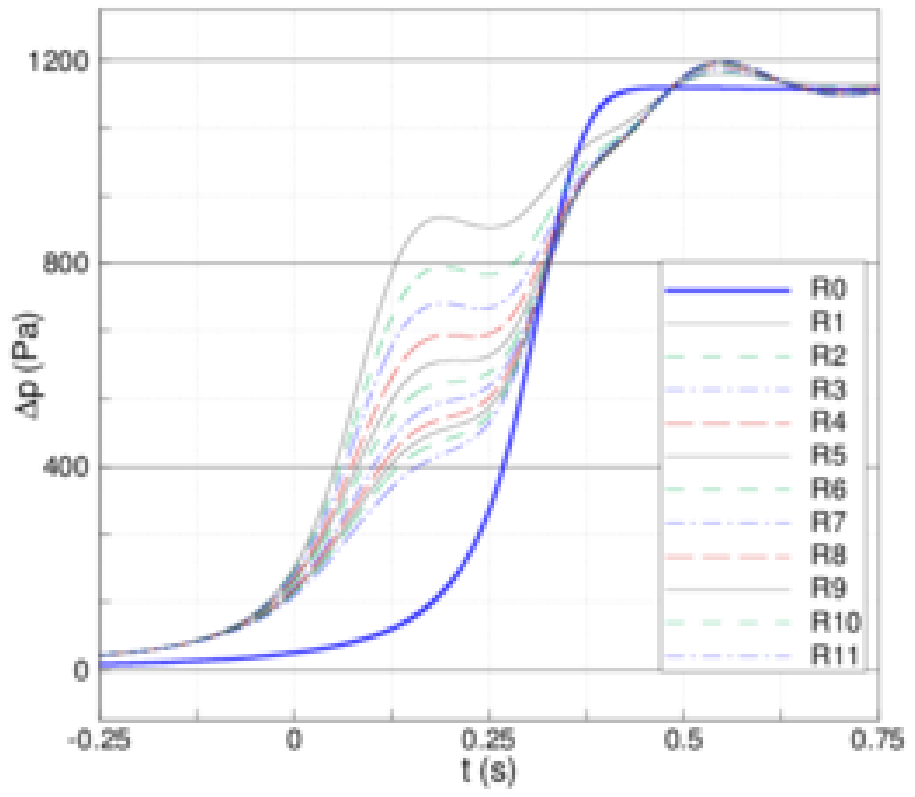}}
\subfigure[pressure gradient]{
\includegraphics[width=0.45\textwidth]{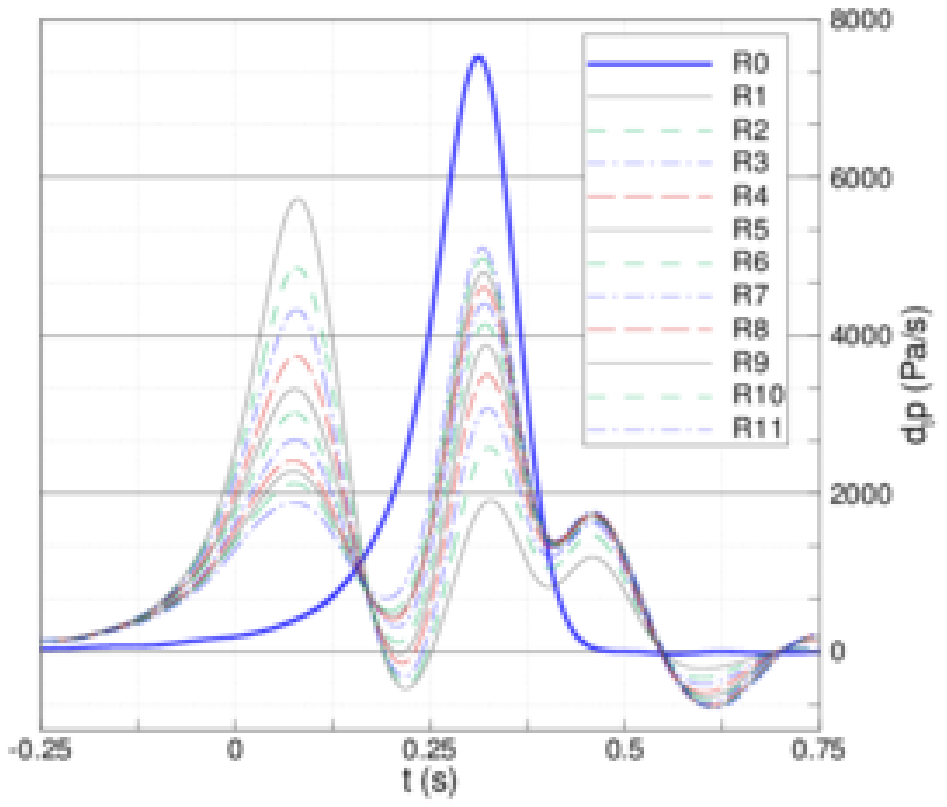}}
\caption{\emph{Temporal history of the pressure and its gradient. Numerical results obtained on the eleven hood sections. $V=250~km/h$, $\sigma=0.152$ and $L_h=20~m$.}}\label{saut_param2}
\end{figure}

\noindent It can be seen that configuration R1, with the smallest hood section, implies the largest first pressure jump. This jump diminishes when the section increases, and it is minimal for the largest section R11. In contrast, the second pressure jump is minimized by configuration R1, and this second pressure jump increases with the hood section becoming maximal for configuration R11. The third pressure jump increases likewise in accordance with the hood section. Indeed, this third pressure jump is caused by the reflection of the rarefaction wave generated, in the hood, by the train's entry in the tunnel. This third jump is, therefore, proportional to the second jump. Graphically, the most efficient hood, with the best "equalization" of both main jumps, is shown to be configuration R4, with a ratio $S_{h}/S_{tun}=1.959$.\\

\begin{table}
\begin{center}
\begin{tabular}{|c|c||c|c||c|c|}
\hline
& $S_{hood}/S_{tunnel}$ & $\Delta p (Pa)$ & diff./R0. (\%) &
$\partial P/\partial t$ max (Pa/s) &  diff./R0. (\%)\\
\hline
R0 & 1 & 1144 & 0 & 7522 &  0\\
 \hline
R1 & 1.397 & 1173 & +2,5 & 5732 & -24 \\
  \hline
R2 & 1.588 & 1176 & +2,8 & 4857 & -35 \\
  \hline
R3 & 1.775 & 1178 & +3 & 4326 & -47 \\
  \hline
R4 & 1.959 & 1188 & +3,8 & 3740 & -50 \\
  \hline
R5 & 2.142 & 1190 & +4 & 3873 & -48 \\
  \hline
R6 & 2.322 & 1192 & +4,2 & 4145 & -45 \\
  \hline
R7 & 2.501 & 1197 & +4,6 & 4407 & -41 \\
  \hline
R8 & 2.678 & 1197 & +4,6 & 4603 & -39 \\
  \hline
R9 & 2.854 & 1197 & +4,6 & 4796 & -36 \\
  \hline
R10 & 3.029 & 1198 & +4,7 & 4978 & -34 \\
  \hline
R11 & 3.203 & 1198 & +4,7 & 5105 & -32 \\
  \hline
\end{tabular}
\end{center}
 \caption{\emph{Pressure magnitude and pressure gradient maxima. Numerical results obtained on the eleven hood sections. $V=250~km/h$, $\sigma=0.152$ and $L_h=20~m$.}}
 \label{max_param2}
\end{table}

\noindent This is confirmed by the values of pressure magnitude and pressure gradient maxima given in Table \ref{max_param2}. Configuration R4 is the most efficient, with a reduction of about 50\% in the temporal pressure gradient. Note that the pressure magnitude increases with the hood section as the third pressure jump increases.\\

\begin{figure}[!h]
\centering \subfigure[global]{
\includegraphics[bb=80 35 665 530,clip=true,scale=.32]{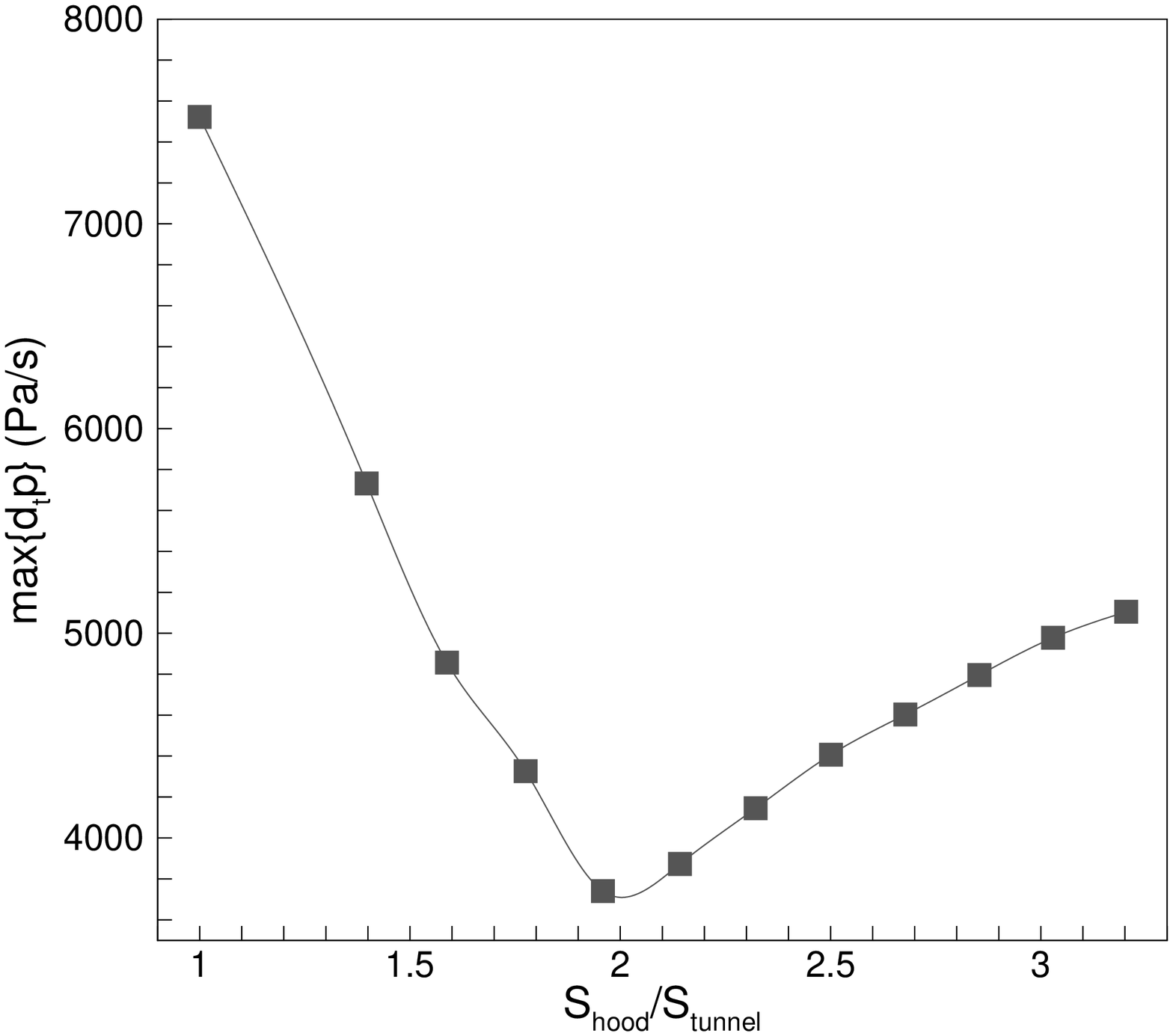}\label{max_grad_param2_1}}
\subfigure[two main jumps]{
\includegraphics[bb=140 35 720 530,clip=true,scale=.32]{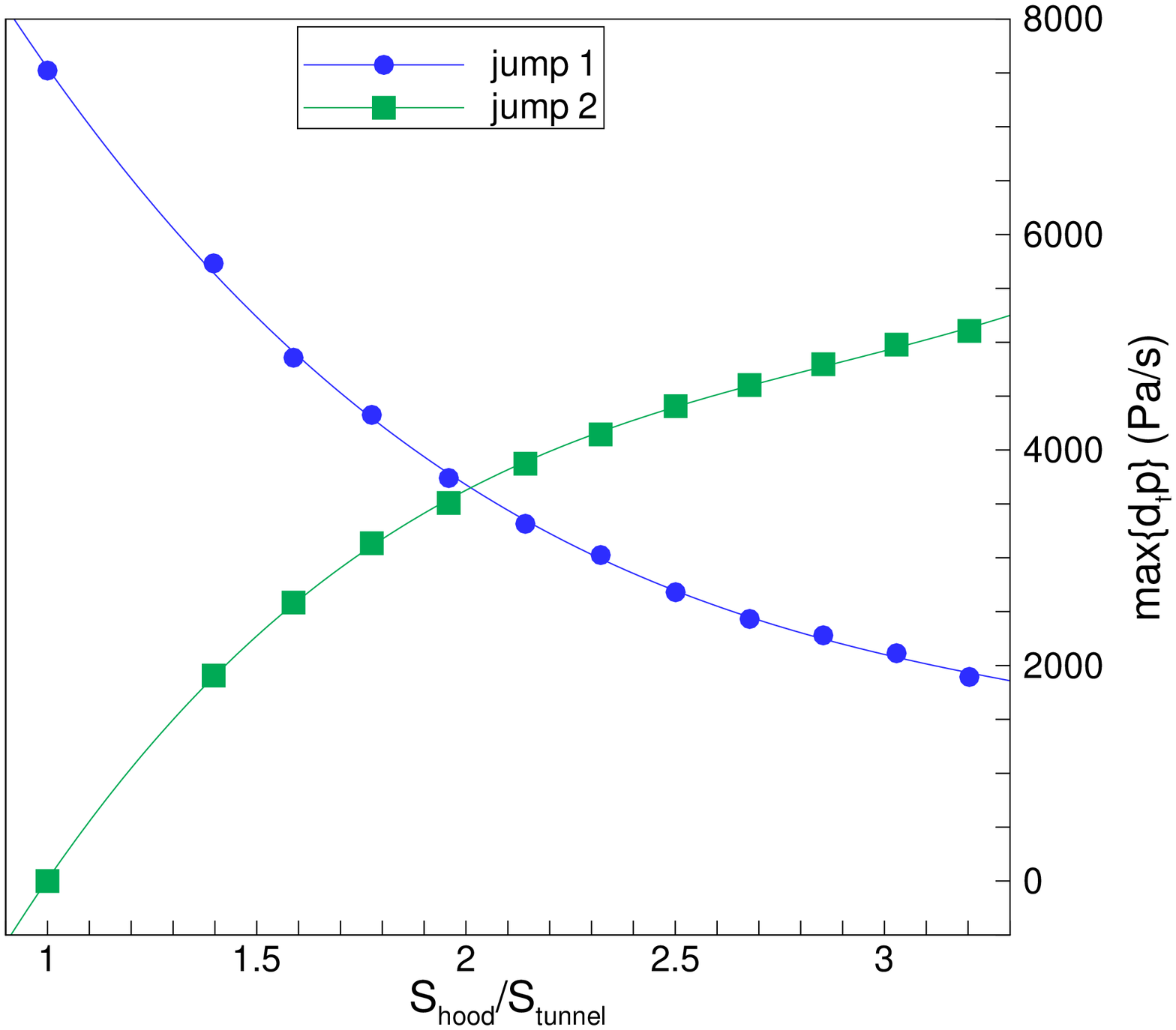}\label{max_grad_param2_2}}
\caption{\emph{Maxima temporal pressure gradient evolution. Numerical results obtained on the eleven hood sections.  $V=250~km/h$, $\sigma=0.152$ and $L_h=20~m$.}}\label{max_grad_param2}
\end{figure}

\noindent Figure \ref{max_grad_param2} shows the evolution of the pressure gradient maxima in figure \ref{max_grad_param2_1} and the evolution of both the main jumps in figure \ref{max_grad_param2_2} with the ratio $S_{h}/S_{tun}$.\\

\noindent The optimal section is given by the minimal value on the first graph \ref{max_grad_param2_1} or by the intersection of both curves on the second graph \ref{max_grad_param2_2}. The section obtained corresponds to the section defined by $S_{h}\simeq2Stun$.

\subsection{Hood length}
\label{ssec:length}

\subsubsection{The test layout}

\noindent In this part, 10 calculations were performed with a hood length dimensioned by the length of the train nose which is an important parameter in the wave generation process. Equation (\ref{length10}) summarizes the ten configurations:
\begin{equation}
\left\{
\begin{array}{l}
\left(L_{h}\right)_{1}=\displaystyle\frac{L_{nose}}{2}=3~m\\
\left(L_{h}\right)_{i}=(i-1)L_{nose},\textmd{  for  }i=2,\ldots,10.
\end{array}
\right.
\label{length10}
\end{equation}
The hood section is the optimal section found in the previous section.

\subsubsection{Results}

\begin{figure}[!h]
\centering \subfigure[pressure]{
\includegraphics[width=0.45\textwidth]{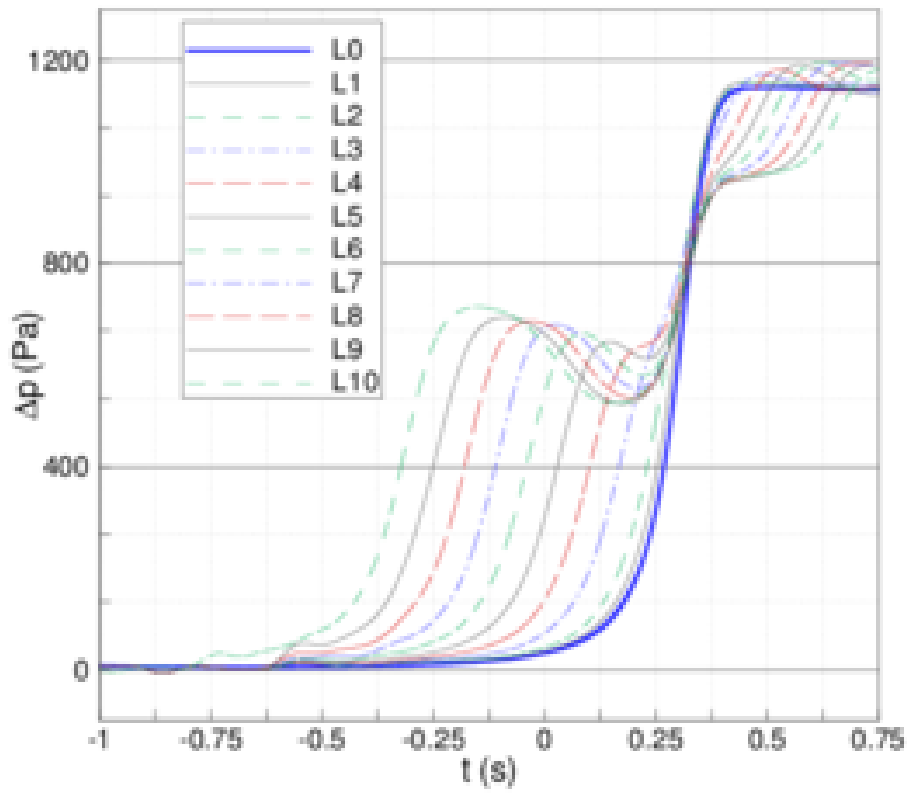}}
\subfigure[pressure gradient]{
\includegraphics[width=0.45\textwidth]{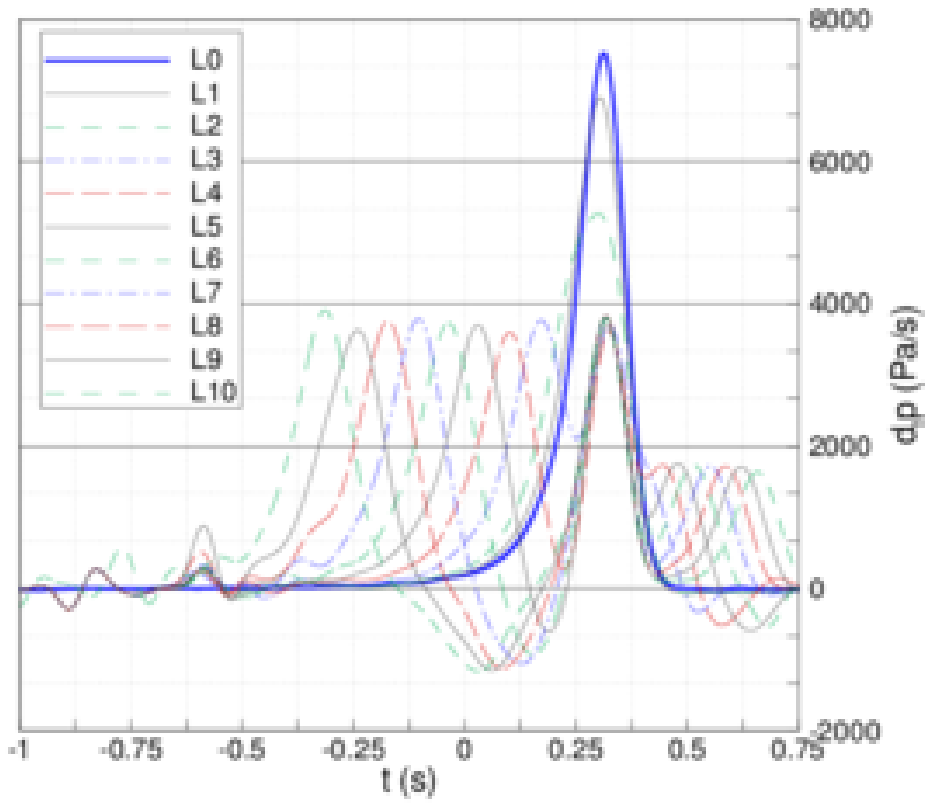}}
\caption{\emph{Temporal history of the pressure and its gradient. Numerical results obtained on the ten hood lengths defined in equation (\ref{length10}). $V=250~km/h$, $\sigma=0.152$ and $S_h/S_{tun}=2$.}}\label{saut_param3}
\end{figure}

Figure \ref{saut_param3} shows the temporal evolution of the pressure, recorded at the numerical sensor, and its gradient. Configuration L0 refers to reference calculation without hood.\\

\noindent As the time $t=0~s$ corresponds to the train's entry in the tunnel, the first pressure jump is shifted in time: the longest hoods generates the earliest jumps and also the most substantial pressure amplitudes. Actually, a long hood leaves more time for wave propagation. Similarly, the rarefaction wave has a greater amplitude when the hood is longer. The second compression wave occurs at the same time for all configurations; indeed, it is generated by the train's entry in the tunnel. The last compression wave, generated by the rarefaction wave induced by the tunnel entry reflection, is delayed in time. The two shortest hoods imply pressure signals near to that of the un-hooded signal. This is confirmed by the pressure gradient. Indeed, the two shortest hoods lead to a lower gain than the others hoods. From configuration L3 to configuration L10, both principle compression waves seem to be equivalent. However, a long hood implies greater fluctuations.\\

\begin{table}
\begin{center}
\begin{tabular}{|c||c|c||c|c|}
\hline
& $\Delta p (Pa)$ & diff./L0. (\%) &
$\partial P/\partial t$ max (Pa/s) &  diff./L0. (\%)\\
\hline
L0 & 1144 & 0 & 7522 &  0\\
 \hline
L1 & 1157 & +1,1 & 6897 & -8 \\
  \hline
L2 & 1157 & +1,1 & 5263 & -30 \\
  \hline
L3 & 1173 & +2,5 & 3772 & -50 \\
  \hline
L4 & 1182 & +3,3 & 3612 & -52 \\
  \hline
L5 & 1191 & +4,1 & 3708 & -51 \\
  \hline
L6 & 1196 & +4,5 & 3791 & -50 \\
  \hline
L7 & 1195 & +4,5 & 3810 & -49 \\
  \hline
L8 & 1194 & +4,4 & 3825 & -49 \\
  \hline
L9 & 1190 & +4 & 3830 & -49 \\
  \hline
L10 & 1184 & +3,5 & 3910 & -48 \\
  \hline
\end{tabular}
\end{center}
 \caption{\emph{Pressure magnitude and pressure gradient maxima. Numerical results obtained on the ten hood lengths defined in equation (\ref{length10}). $V=250~km/h$, $\sigma=0.152$ and $S_h/S_{tun}=2$.}}
 \label{max_param3}
\end{table}

\noindent Table \ref{max_param3} summarizes the maxima of pressure and pressure gradients. It shows that the optimal hood is L4 (3 times the train's nose length) configuration which leads to 3612~Pa/s. This hood induced an optimal crossing of the rarefaction wave between both main compression waves. A substantial range, between configurations L3 (2 times the train's nose length) and L9 (8 times the train's nose length) leads to a gain approaching 50\%. Therefore, a common hood length can be considered such that the hood remains optimal for different trains and for different velocities.\\

\begin{figure}[!h]
\centering \subfigure[global]{
\includegraphics[bb=80 35 665 530,clip=true,scale=.32]{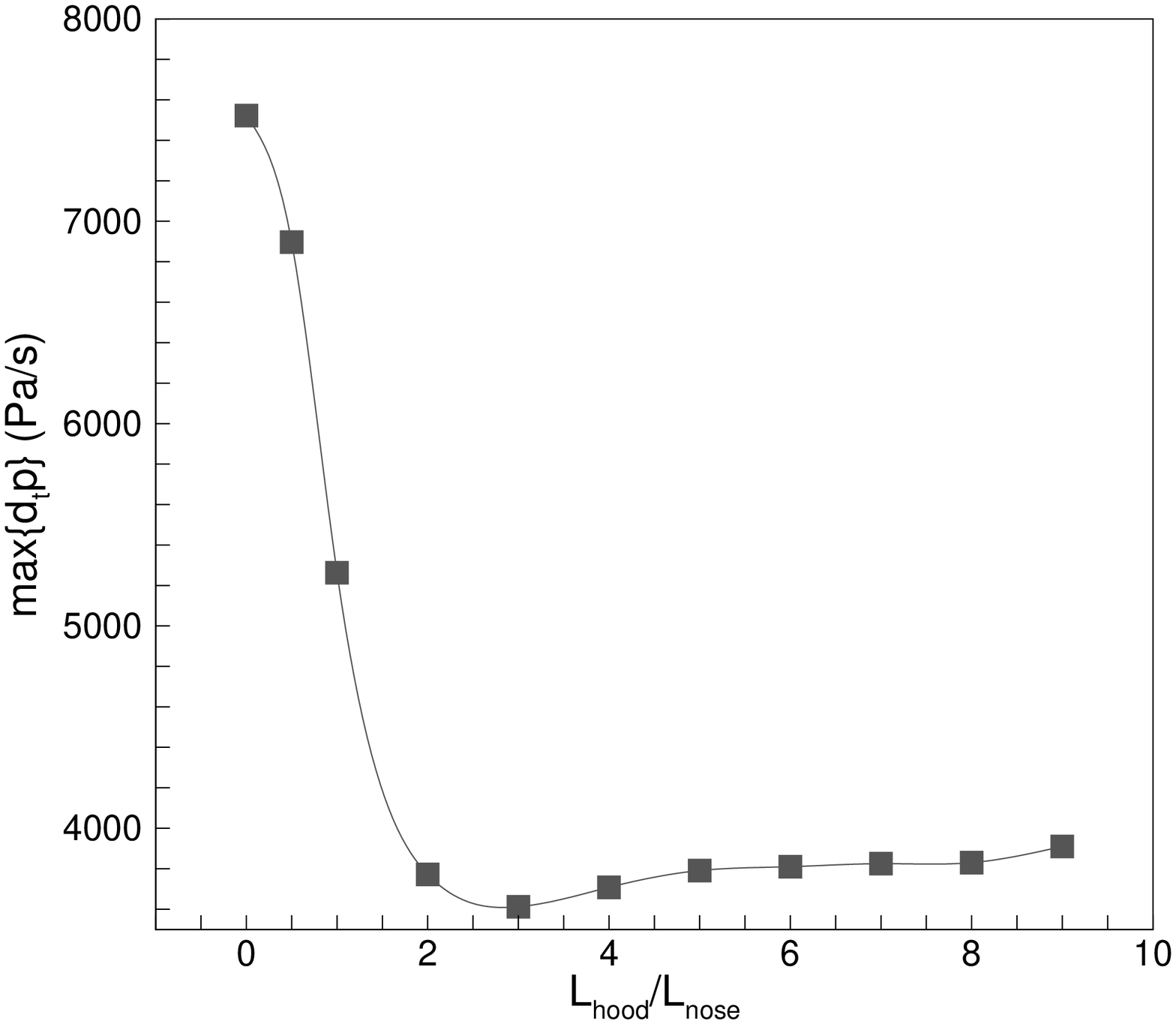}\label{max_grad_param3_1}}
\subfigure[both main jumps]{
\includegraphics[bb=140 35 720 530,clip=true,scale=.32]{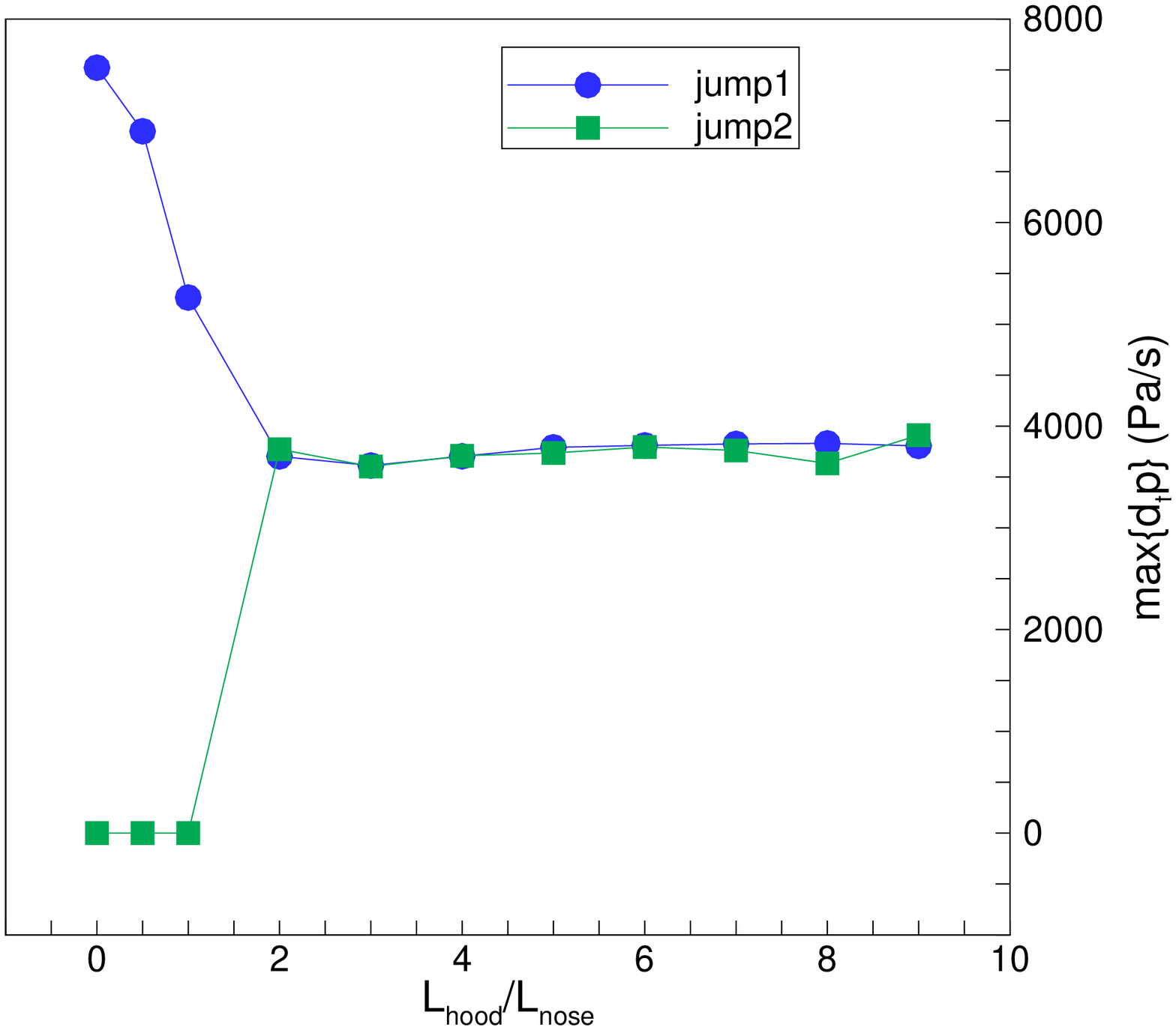}\label{max_grad_param3_2}}
\caption{\emph{Maxima temporal pressure gradient evolution. Numerical results obtained on the ten hood lengths defined in equation \ref{length10}. $V=250~km/h$, $\sigma=0.152$ and $S_h/S_{tun}=1.775$.}}\label{max_grad_param3}
\end{figure}

\noindent The graphs in figure \ref{max_grad_param3} represent the evolution of the pressure gradient maxima \ref{max_grad_param3_1} and the evolution of both main pressure gradient maxima \ref{max_grad_param3_2} versus the non-dimensional hood length $L_{h}/L_{nose}$. The range [L3;L9], in which the gain is constant, is clearly highlighted. Configuration L10 generates an increase in the pressure gradient maxima : if the hood is too long, it behaves like a tunnel (steepening of waves).\\

\noindent The effect of the hood section is clearly more important than its length, and it is possible to take a common hood length for different velocities. Indeed, the Mach number of the train plays an important role in the compression wave generation process. It is interesting to study the effect of the train's velocity on the optimal hood section.

\subsection{Effect of the train's velocity on the optimal hood section}
\label{ssec:velocity}

As said previously, the hood length is not primordial. For a velocity of 250~km/h, a hood length of between 12~m and 48~m gives approximately the same result. This range certainly evolves with the train's velocity, but it can be supposed that a hood of 20~m in length is in the ideal range for higher velocities. In this study, the hood section is only considered for velocities of 275~km/h, 300~km/h, 325~km/h, 350~km/h and finally 400~km/h. The first two correspond to current high-speed train velocities. With the last three, the rarefaction wave $a_{2r}^{-}$ of the figure \ref{diag_wave} does not pass between both main compression waves. Indeed, if we denote by $t_{0}$ the time at which the train enters the hood, the time at which the rarefaction wave reaches the tunnel entry is $t_{0}+(3L_{h}+d)/c$, where $d$ is the hydraulic diameter and $c$ the speed of sound. Whereas the time at which the second compression wave is generated in the tunnel, corresponding to the train's entry, is $t_{0}+L_{h}/Mc$, where $M$ is the train's Mach number. It can easily be seen that when the train's Mach number is over 1/3, the train reaches the tunnel entry before the rarefaction wave. In this way, it is possible to see the effect of the rarefaction wave on the optimal hood section.\\

\begin{figure}[!h]
\centering \subfigure[vs. section ratio]{
\includegraphics[bb=85 35 675 535,clip=true,scale=.32]{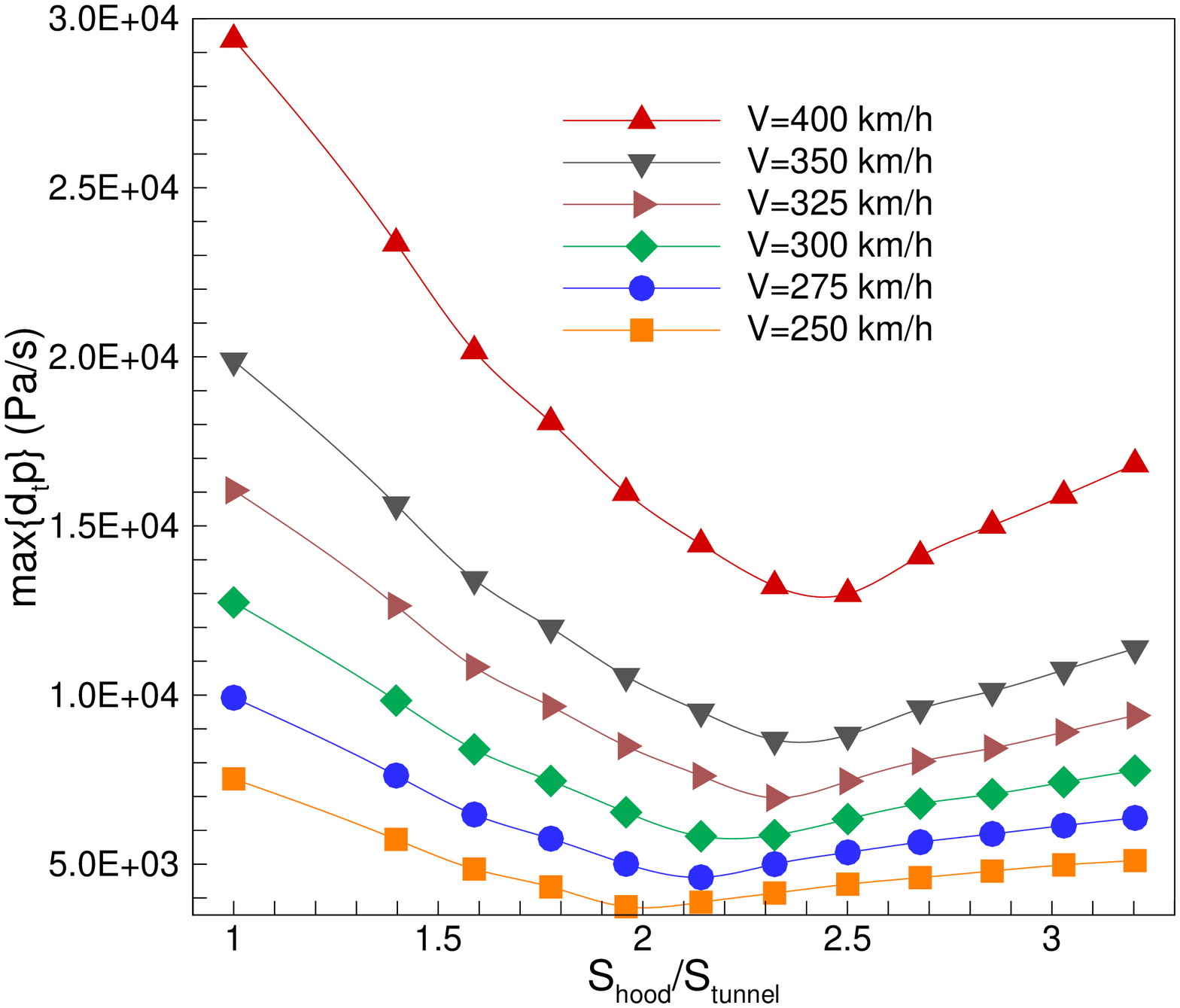}}
\subfigure[vs. Mach cube]{
\includegraphics[bb=140 35 720 535,clip=true,scale=.32]{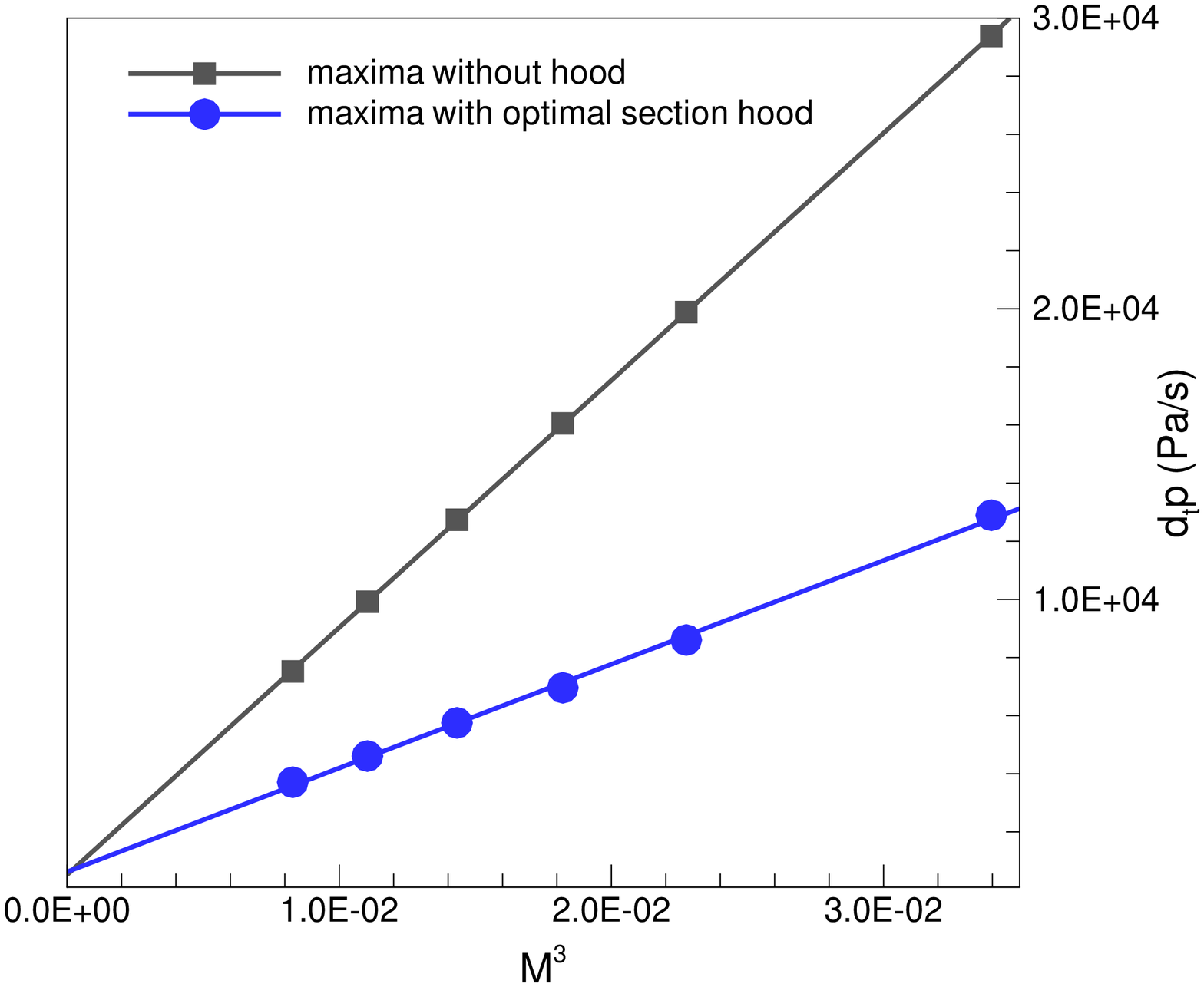}}
\caption{\emph{Gradient maxima evolution. Numerical results obtained on the eleven hood sections for the six velocities 250~km/h, 275~km/h, 300~km/h, 325~km/h, 350~km/h and 400~km/h. $\sigma=0.152$ and $L_h=20~m$.}}\label{saut_param6}
\end{figure}

\noindent Figure \ref{saut_param6} shows the results. The graph on the left represents the evolution of the pressure gradient maxima versus the ratio hood section and tunnel section for all velocities. And the graph on the right is the evolution of the pressure gradient maxima versus the Mach number cube, for the configuration without hood and for the optimal hood section.\\

\noindent The graph on the left shows that the optimal ratio $S_{h}/S_{tun}$ increases with the velocity. However, for the lowest three velocities, corresponding to current velocities, the ratio evolution is weak. The difference between the optimal ratio at $V=250~km/h$ and $V=300~km/h$ is only 10 per cent.\\
Evolutions of both gradient maxima (without hood and with optimal hood section) versus the cube of the Mach number, graph on the right, are linear. This is a well-known result without a hood, see equation (\ref{grad_max_formula}), and as is shown here the maximum of the temporal pressure gradient for the optimal hood can be known by the train's Mach number.\\

\begin{figure}[!h]
\centering
\includegraphics[bb=80 235 650 565,clip=true,scale=.5]{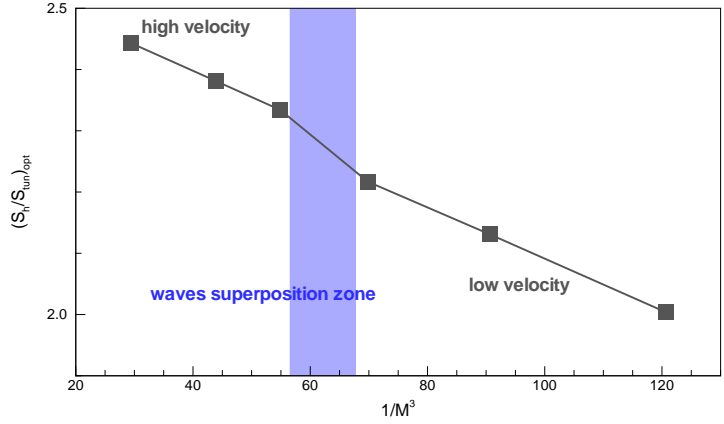}
\caption{\emph{Optimal section hood evolution versus the Mach number cube inverse. Numerical results obtained on the eleven hood sections for the six velocities 250~km/h, 275~km/h, 300~km/h, 325~km/h, 350~km/h and 400~km/h.  $\sigma=0.152$ and $L_h=20~m$.}}\label{rap_opt}
\end{figure}

\noindent Figure \ref{rap_opt} shows the evolution of the optimal hood section versus the Mach number cube inverse.\\

\noindent The fact that the optimal section increases with the train's velocity is highlighted. The curve is clearly divided into two linear zones. The first one corresponds to the low velocities, where the rarefaction wave is located between the two main compression waves. The second one is obtained with high velocities, for which the rarefaction wave is behind the second compression wave. The slopes of these two lines are the same. Between these two lines, the sudden jump is recorded when the rarefaction wave and the second compression wave are superimposed.

\section{Conclusion}

The effects of tunnel hood on the maximum pressure and the temporal gradient of the pressure were investigated by the help of a well-tested three-dimensional numerical tools. The parametric study dealt with the shape, the cross section and the length of the hoods.
\begin{itemize}
   \item The comparison of the use respectively of an elliptic, conical and constant evolution of the hood shape, showed that the last configuration was the most efficient. Indeed, a reduction by 46\% of the maximum pressure gradient was obtained.
   \item The effects of the cross section of the hoods showed that the optimal ratio R between the hood section and the tunnel section was about 2: the maximum pressure gradient was reduced by 50\% with a R equal to 1.959.  
   \item The influence of the length was characterized by using the ratio between the hood length and the nose length. It was shown that when this ratio increases the maximum pressure gradient decreases and this up to R=2 where the reduction was about 50\%. This reduction remained quasi-constant until R=8. Beyond this value, the maximum of pressure gradient increased. This means that when the hood length is greater than 8 times the train nose length, the hood behaves like a tunnel.
\end{itemize}

\noindent 

\noindent Finally, it was shown that the optimal section increases with the train's velocity. It was, in particular, shown that the rarefaction wave $a_{2r}^{-}$ plays an important role in the determination of the optimal hood section. \\

\noindent A perspective work could be to study the effects of perforated hoods. Indeed, perforations modify the initial compression wave and, hence, could reduced further the disturbances.

\bibliographystyle{elsarticle-num}
\bibliography{tout_TL}

\begin{thebibliography}{10}
\expandafter\ifx\csname url\endcsname\relax
  \def\url#1{\texttt{#1}}\fi
\expandafter\ifx\csname urlprefix\endcsname\relax\def\urlprefix{URL }\fi
\expandafter\ifx\csname href\endcsname\relax
  \def\href#1#2{#2} \def\path#1{#1}\fi

\bibitem{mae93}
T.~Maeda, T.~Matsumura, M.~Iida, K.~Uchida, Effect of shape of train nose on
  compression wave generated by train entering tunnel, Proceedings of the
  International Conference on Speedup Technology for Railway and Maglev
  Vehicles 2 (1993) 315--319.

\bibitem{oga_fuj97}
T.~Ogawa, K.~Fujii, Numerical investigation of three-dimensional compressible
  flows induced by a train moving into a tunnel, Computers \& Fluids 26 (1997)
  565--585.

\bibitem{bel_kag02}
M.~Bellenoue, T.~Kageyama, Train/tunnel geometry effects on the compression
  wave generated by a high-speed train, Notes on Numerical Fluid Mechanics and
  Multidisciplinary Design 79 (2002) 276--289.

\bibitem{ku10}
Y.~Ku, J.~Rho, S.~Yun, M.~Kwak, K.~Kim, H.~Kwon, D.~Lee, Optimal
  cross-sectional area distribution of a high-speed train nose to minimize the
  tunnel micro-pressure wave, Struct. Multidisc. Optim. 42 (2010) 965--976.

\bibitem{ki11}
K.~Kikuchi, M.~Iida, T.~Fukuda, Optimization of train nose shape for reducing
  micro-pressure wave radiated from tunnel exit, Journal of Low Frequency
  Noise, Vibration and Active Control 30 (2011) 1--19.

\bibitem{how99}
M.~S. Howe, On the compression wave generated when a high speed train enters a
  tunnel with a flared portal, Journal of Fluids and Structures 13 (1999)
  481--498.

\bibitem{ret_gre02_2}
J.-M. R\'ety, R.~Gr\'egoire, Numerical investigation of tunnels extensions
  attenuating the pressure gradient generated by a train entering a tunnel,
  Notes on numerical fluid mechanics and multidisciplinary design 79 (2002)
  239--248.

\bibitem{how06}
M.~S. Howe, M.~Ida, T.~Maeda, Y.~Sakuma, Rapid calculation of the compression
  wave generated by a train entering with a vented hood, Journal of Sound and
  Vibration 297 (2006) 267--292.

\bibitem{xia10}
X.~Xiang, L.~Xue, Tunnel hoods effects on high speed train-tunnel compression
  wave, Journal of Hydrodynamics 22(5) (2010) 897--904.

\bibitem{liu10}
T.~Liu, H.~Tian, X.~Liang, Design and optimization of tunnel hoods, Tunnel and
  Underground Space technology 25 (2010) 212--219.

\bibitem{Heine12}
D.~Heine, K.~Ehrenfried, Experimental study of the compression-wave generation
  due to train-tunnel entry, Proceedings of the First International Conference
  on Railway Technology: Research, Development and Maintenance, Paper 163
  (2012).

\bibitem{li11}
X.~Li, D.~Chen, F.~Xie, Y.~Z, Unsteady simulation for a high-speed train
  entering a tunnel, Journal of Zhejiang University-Science A 12 (2011)
  957--963.

\bibitem{ko12}
Y.~Ko, C.~Chen, I.~Hoe, S.~Wang, Field measurements of aerodynamic pressures in
  tunnels induced by high speed trains, Journal of Wind Engineering and
  Industrial Aerodynamics 100 (2012) 19--29.

\bibitem{roe81}
P.~Roe, Approximate riemann solvers, parameter vectors, and difference schemes,
  Journal of Computational Physics 43 (1981) 357--372.

\bibitem{van79}
B.~Van~Leer, Towards the ultimate conservative difference scheme v.a. second
  order sequel to godunov's method, SIAM J. Sci. Stat. Comput. 32 (1979)
  101--136.

\bibitem{bar_jes89}
T.~Barth, D.~Jespersen, The design and application of upwind schemes on
  unstructured meshes, AIAA Paper 89-0366 (1989).

\bibitem{uys11}
D.~Uystepruyst, M.~William-Louis, E.~Creus\'e, S.~Nicaise, F.~Monnoyer,
  Efficient 3{D} numerical prediction of the pressure wave generated by
  high-speed trains entering tunnels, Computers \& Fluids, 47 (2011) 165--177.

\bibitem{ret_gre02_1}
J.-M. R\'ety, R.~Gr\'egoire, Numerical simulation of the pressure wave
  generated by a train enters a tunnel, TRANSAERO - A European initiative on
  transient aerodynamics for railway system optimisation, results of the
  Brite/Euram project Transient aerodynamics for railway system optimisation
  (2002) 225--238.

\bibitem{woo_pop92}
C.~W. Pope, W.~A. Woods, Boundary conditions for the transit of a train through
  a tunnel with special reference to the entry and exit mesh fraction and the
  contact surface, Aerodynamics and Ventilation of Vehicles Tunnel (1992)
  79--105.

\bibitem{mwl_tou05}
M.~J.-P. William-Louis, C.~Tournier, A wave signature based for the prediction
  of pressure transients in railway tunnels, J. of wind engeneering and
  industrial aerodynamics 93 (2005) 521--531.

\bibitem{oza_mae88}
S.~Ozawa, T.~Maeda, Tunnel entrance hoods for reduction of micro-pressure wave,
  Quarterly Report of Railway Technical Research Institute 29 (1988).

\bibitem{bel01}
M.~Bellenoue, B.~Auvity, T.~Kageyama, Blind hood effects on the compression
  wave generated by a train entering a tunnel, Experimental Thermal and Fluid
  Science 25 (2001) 397--407.

\bibitem{oga_fuj96}
T.~Ogawa, K.~Fujii, Prediction and allievation of a booming noise created by a
  high-speed train moving into a tunnel, Computational Fluids Dynamics (1996)
  808--814.

\bibitem{mwl_tou03}
M.~J.-P. William-Louis, C.~Tournier, Numerical and experimental study of
  transversal pressure waves at a tube exit, Experimental Thermal ans Fluid
  Science 28 (2003) 525--532.

\bibitem{iid01}
M.~Iida, Y.~Tanaka, K.~Kikuchi, T.~Fukuda, Pressure waves radiated directly
  from tunnel portals at train entry or exit, Quarterly report of railway
  technical research institute 42, 2 (2001) 83--88.

\end{thebibliography}

\end{document}